\documentclass[twocolumn,10pt]{IEEEtran}
\normalsize

\usepackage{enumerate}
\usepackage{color}

\usepackage{cite}

\usepackage{graphicx}
\ifCLASSINFOpdf
\else
\fi

\usepackage{amsthm}
\usepackage{amssymb,amsfonts}
\usepackage{amsfonts,balance}

\usepackage[cmex10]{amsmath}

\newtheorem{theorem}{Theorem}

\usepackage{algorithmic}

\usepackage{array}
\usepackage{multirow,makecell}
\usepackage[caption=false,font=footnotesize]{subfig}

\usepackage{verbatim}

\begin{document}
%
\title{Digital Signal Processing for Molecular Communication via Chemical Reactions-based Microfluidic Circuits}

%
%
%

\author{Dadi~Bi,~\IEEEmembership{Student Member,~IEEE,}
        Yansha~Deng,~\IEEEmembership{Member,~IEEE,}
\thanks{D. Bi and Y. Deng are with the Department of Engineering, King’s College London, London WC2R 2LS, U.K. (e-mail:\{dadi.bi, yansha.deng\}@kcl.ac.uk). (Corresponding author: Yansha Deng).}
}
\maketitle

\begin{abstract}
Chemical reactions-based microfluidic circuits are expected to provide new opportunities to perform signal processing functions over molecular domain. To realize this vision, 
in this article, we exploit and present the digital signal processing capabilities of chemical reactions-based microfluidic circuits. Aiming to facilitate microfluidic circuit design, we describe a microfluidic circuit using a five-level architecture: 1) Molecular Propagation; 2) Chemical Transformation; 3) Microfluidic Modules; 4) Microfluidic Logic Gates; and 5) Microfluidic Circuits. We first identify the components at Levels 1 and 2, and present how their combinations can build the basic modules for Level 3. We then assemble basic modules to construct five types of logic gate for Level 4, including AND, NAND, OR, NOR, and XOR gates, which show advantages of microfluidic circuits in reusability and modularity. Last but not least, we discuss challenges and potential solutions for designing, building, and testing microfluidic circuits with complex signal processing functions in Level 5 based on the digital logic gates at Level 4.

\end{abstract}


%
\IEEEpeerreviewmaketitle

\section{Introduction}
Molecular communication (MC) employs chemical signals to exchange information and is considered as a promising methodology to facilitate a diversity of applications, ranging from personalized healthcare to manufacturing industry. 
Although the literature on the design of MC systems has grown considerably over the past few years, existing work has been mostly theoretical in nature, and functioning MC testbed implementation is few. In particular, \cite{farsad2013tabletop,giannoukos2018chemical,8924625} have developed functioning MC prototypes, where transmitted bit sequence was modulated to concentrations of alcohol molecules \cite{farsad2013tabletop}, odor molecules \cite{giannoukos2018chemical}, and protons  \cite{8924625}. It is noted that their signal processing functions on chemical concentration signals are all achieved via external electronic devices, such as electric spray, odor emitter, and Arduino controlled LED. However, the utilization of electronic devices can hardly meet the biocompatible, non-invasive, and size-miniaturized requirements of biomedical-related applications.

The above limitations of electronic devices inspire us to design novel MC devices to perform signal processing functions directly over chemical signals rather than electrical signals. In general, signal processing functions can be realized over molecular domain in a twofold fashion, namely, 1) genetic circuits \cite{weiss2003genetic} in engineered living cells, and 2) chemical circuits \cite{cook2009programmability} based on ``non-living'' chemical reactions. Genetic circuits engineer cell behaviors by embedding synthetic gene networks to produce desired responses. 
While genetic circuits offer biocompatibility, non-invasiveness, and miniaturization, they still currently face challenges, such as slow speed, unreliability, and nonscalability, which motivate us to use chemical circuits as an alternative approach. Chemical circuits execute signal processing functions by means of chemical reaction networks (CRNs). A CRN is defined as a finite set of chemical reactions comprising a finite number of species. These reactions occur in a well-stirred environment to realize a function or an algorithm via mass action kinetics.

The CRNs can integrate with microfluidic systems to construct chemical reactions-based microfluidic circuits. A microfluidic system processes or manipulates small amount of fluids using channels at dimensions of tens to hundreds of micrometers \cite{bruus2008theoretical}. Therefore, chemical reactions-based microfluidic circuits are not only endowed with advantages of rapid analysis and low reagent costs due to size reduction, but can also benefit from an additional space level of chemical control through applying and regulating chemical reactions in different regions of a microfluidic device.

\begin{figure}%
	\centering
	\includegraphics[width=3in]{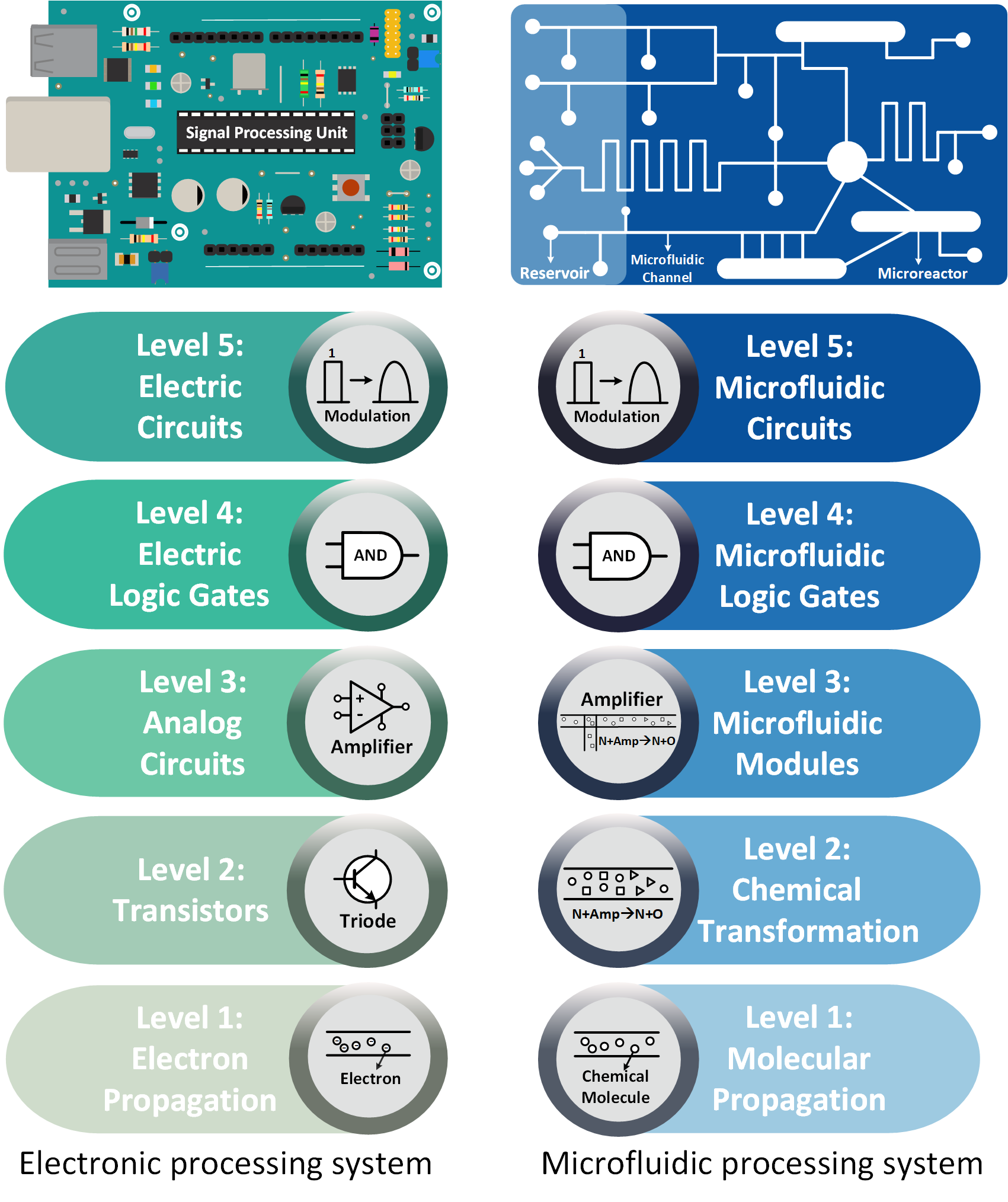}
	\caption[]{The illustration of levels of abstraction for an electronic processing system and a microfluidic processing system along with typical building components at each level.}
	\label{f_vision}%
\end{figure}

The concept of using chemical reactions-based microfluidic circuits to execute signal processing functions was first introduced and investigated in \cite{8255057,9090868}, where an MC transceiver was designed to successfully realize binary concentration shift keying (CSK) modulation and demodulation functions. Knowing the advantages of microfluidic circuits, we envision the utilization of microfluidic circuits to perform more complex signal processing functions for MC.

In nature, many cellular signaling processes can be interpreted to be driven by digital signals with two discrete states according to whether signals are stronger than thresholds. A typical example is the bacterial coordination behavior after the detection of a sufficiently high autoinducer concentration representing bit-1 from a low autoinducer concentration representing bit-0.
In effect, digital signals are ideal for reliable state transitions and signal integration, and are useful for decision-making circuits \cite{xiang2018scaling}. Furthermore, digital circuits can be easily scalable and
are popular in wireless signal processing. 
Motivated by these facts, in this article, we exploit the ability of chemical reactions-based microfluidic circuits to process digital chemical signals. The contributions of this article are summarized as follows:
\begin{itemize}
	\item To facilitate digital microfluidic circuit design, we propose a five-level architecture to describe a microfluidic circuit with a discussion of the components in each level. 
	\item We present the microfluidic designs of AND, NAND, OR, NOR, and XOR logic gates, which are then validated by simulation results.
	\item Finally, we illustrate a roadmap for the development of microfluidic circuits with complex signal processing functions. Importantly, we also discuss the challenges during circuit design and testing, and provide corresponding potential solutions.
\end{itemize}

The remainder of this article is organized as follows. First, we present the five-level architecture of microfluidic circuits and discuss the components at Levels 1--3 in Section \ref{b4}. Then, we detail the microfluidic designs of logic gates for Level 4 in Section \ref{digital_gates}. Finally, we identify challenges and potential solutions to build large-scale microfluidic circuits for Level 5 in Section \ref{challenge}.


\section{Microfluidic Circuit Architecture and Module Design}
\label{b4}
\begin{figure*}[!t]
	\centering
	\includegraphics[width=7.15in]{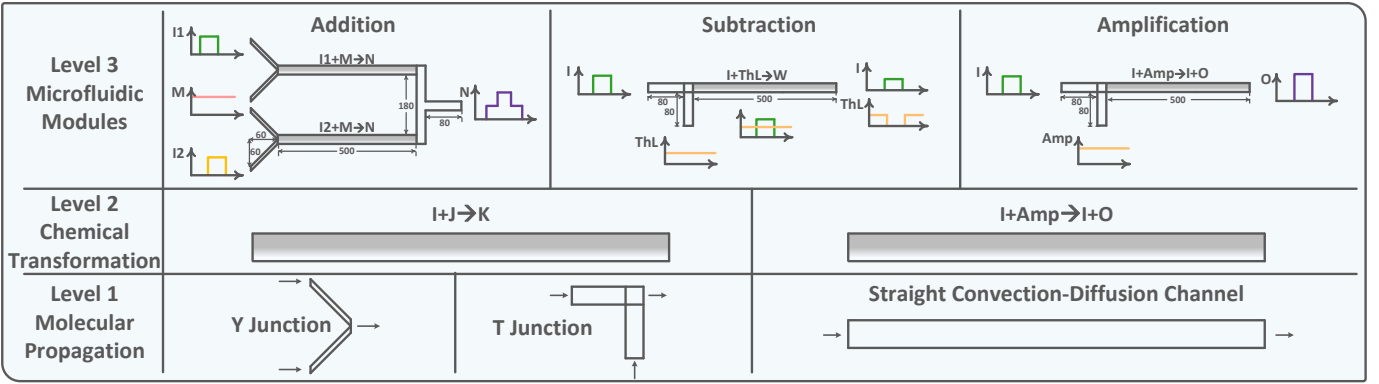}
	\caption{The components at Levels 1 and 2, and their construction to the addition, subtraction, and amplification modules for Level 3. The unit of the channel length is $\mu$m.}
	\label{basic_modules}
\end{figure*}

In order to facilitate microfluidic circuit design, we abstract a microfluidic processing system into five levels as shown in Fig. \ref{f_vision}. The levels of the abstraction are as follows:
\begin{itemize}
	\item \textbf{Level 1: Molecular Propagation} -- the movement of chemical molecules in microfluidic channels. 
	\item \textbf{Level 2: Chemical Transformation} -- the interaction between different species, i.e., the chemical reactions that support various signal processing functions.
	\item \textbf{Level 3: Microfluidic Modules} -- the basic modules performing simple calculations.
	\item \textbf{Level 4: Microfluidic Logic Gates} -- the digital logic gates assembled from the microfluidic modules designed for Level 3. Although the microfluidic modules at Level 3 process a continuous range of concentrations, an appropriate combination of them can lead to digital signal operation.
	\item \textbf{Level 5: Microfluidic Circuits} -- the top level is microfluidic circuit itself, which is built from the logic gates designed for Level 4 and can perform a specific signal processing function, such as coding-decoding and modulation-demodulation for MC.
\end{itemize}

As shown in Fig. \ref{f_vision}, we map our proposed levels of microfluidic processing systems to those of electronic processing systems \cite{harris2010digital}. The main differences lie in Levels 1 and 2. In electronic systems, Level 1 focuses on propagation of electrons. By contrast, Level 1 in microfluidic systems is based on the movement of chemical particles. 
The Level 2 of electronic systems is composed of transistors (e.g., diode and triode), whereas that of microfluidic systems is based on chemical reactions for signal transformation. The Levels 3--5 in these two systems have similar functions, but are realized differently via either electronic components or microfluidic components. In the following, we present the components at Levels 1 and 2, and then focus on how these components support the construction of the microfluidic modules at Level 3.

\subsection{Level 1: Molecular Propagation}

In Level 1, the movement of chemical molecules is bounded by channel geometry, and its dispersion is governed by diffusion 
and convection. 
A channel without reactions refers to a convection-diffusion channel. According to the channel shape, we consider three types of geometry: Y junction, T junction, and straight convection-diffusion channel as shown in Fig. \ref{basic_modules}. The Y junction and T junction are both merging channels with two inlets and one outlet, and they can facilitate the mixing of different species injected through two inlets.
The straight convection-diffusion channels only provide a pathway for chemical molecules.

\subsection{Level 2: Chemical Transformation}
In Level 2, we introduce chemical reactions into convection-diffusion channels. The channels with chemical reactions are named as convection-diffusion-reaction channels and are filled with grey-gradient color as shown in Fig. \ref{basic_modules}. In this article, we consider two forms of chemical reactions: $I+J\to K$ and $I+Amp\to I+O$ \cite{scalise2014designing}. By introducing chemical reactions, microfluidic circuits are endowed with signal processing capability.




\subsection{Level 3: Microfluidic Modules}
By combining the components at Levels 1 and 2, we can construct the \textit{Addition}, \textit{Subtraction}, and \textit{Amplification} modules for Level 3. As illustrated in Fig. \ref{basic_modules}, each module contains one or two chemical reactions. In the following, we reveal the signal transformation nature of the reactions at Level 2 and discuss the mechanism of each module.

\subsubsection{Addition Module} 
The addition module calculates the total concentration of two different molecular species and is achieved by converting them to the same molecular species. 
As shown in Fig. \ref{basic_modules}, it is composed of two Y junctions, two reaction channels, and convection-diffusion channels. 
The inputs of an addition module are three chemical signals containing species $I1$, $M$, and $I2$, respectively. The output is the chemical signal with species $N$.
In reaction channels, species $I1$ and $I2$ are transformed to species $N$ via $I1+M\to N$ and $I2+M\to N$. Due to the one-to-one stoichiometric relationship between reactants and product, the generated concentration of species $N$ equals the consumed concentration of species $I1/I2$. Moreover, the stoichiometric relationship also reveals that the amount of transformed species $I1/I2$ depends on the concentration of species $M$. To ensure a complete conversion of species $I1/I2$ to species $N$, the concentration of species $M$ needs to be greater than or at least equal to the concentration of species $I1/I2$. After reactions, the species $N$ generated in two reaction channels converge at convection-diffusion channels to generate the final output.


\subsubsection{Subtraction Module}
The subtraction module calculates the concentration difference between two species and relies on the depletion of one species by the other species.
As shown in Fig. \ref{basic_modules}, it is consisted of a T junction and a reaction channel with species $I$ and $ThL$ as inputs and the remaining species $I$ as output. 
In the reaction channel, input species $I$ is consumed by species $ThL$ via $I+ThL\to W$, where species $W$ represents a waste species whose concentration we do not keep track of. The module output, i.e., the remaining concentration of species $I$, is determined by the concentration of species $ThL$. Under the condition that the concentration of species $ThL$ is greater than the concentration of species $I$, species $I$ will be fully depleted.
As a result, the concentration of species $I$ will be set as zero due to the one-to-one stoichiometric relationship between species $I$ and species $ThL$. 

\subsubsection{Amplification Module} 
The amplification module generates a chemical signal whose width and amplitude are determined by two input signals.
As shown in Fig. \ref{basic_modules}, it uses the same geometry structure as the subtraction module but with a different reaction $I+Amp\to I+O$.   
In the presence of species $I$, species $I$ acts as a catalyst to enable the conversion of the other input species $Amp$ to output species $O$; if species $I$ is absent, species $O$ will not be produced. In this way, species $I$ determines the time period when output species $O$ is generated, whereas reactant $Amp$ influences the concentration of species $O$. The higher the concentration of species $Amp$, the higher the concentration of species $O$, which allows us to flexibly adjust the output concentration of species $O$. 

\section{Level 4: Microfluidic Logic Gates}
\label{digital_gates}

Flowing fluids in microfluidic channels allows for an easy serial processing operation, which endows microfluidic circuits with the feature of integrating  different functional modules to build the microfluidic logic gates at Level 4. In particular, we apply the microfluidic modules designed for Level 3 to construct the AND, NAND, OR, NOR, and XOR gates. Throughout this paper, the HIGH state (bit-1) and the LOW state (bit-0) are represented by non-zero concentration and zero concentration, respectively.

\subsection{AND and NAND Gates}

We first design the AND gate as shown in Fig. \ref{f_4}, which consists of the addition, subtraction, and amplification modules.
The AND gate takes input signals $I1$ and $I2$ and produces a HIGH state for output species $O$ only when both inputs are HIGH.
The addition module first converts species $I1$ and $I2$ to an intermediate species $N$ assisted by species $M$. According to the combination of input species $I1$ and $I2$, the species $N$ concentration $C_N$ at location $x_1$ has a ladder-shaped distribution (the purple line in the subtraction module in Fig. \ref{f_4}) with three typical values: 
\begin{itemize}
	\item $C_N=0$ when both input species are LOW,
	\item $C_N=\alpha_1$ when only one input species is HIGH (the red dot line in Fig. \ref{f_4}),
	\item $C_N=\alpha_2$ when both input species are HIGH (the red dash line in Fig. \ref{f_4}).
\end{itemize}  
To achieve an AND function, the concentration of species $N$ in the amplification module is required to span over the time period where both species $I1$ and $I2$ are HIGH. Therefore, the species $N$ generated by the addition module flows into a subtraction module and undergoes a depletion by species $ThL$ that is continuously supplied through the first T junction. We highlight that the concentration of species $ThL$ at location $x_1$, i.e., $C_{ThL}$, must satisfy $\alpha_1<C_{ThL}<\alpha_2$ so that the remaining concentration of species $N$ is larger than zero only when both inputs are HIGH. 
Once the remaining species $N$ arrives at the amplification module, reaction $N+Amp\to N+O$ is activated, inducing the conversion of species $Amp$ to output species $O$. Thus, we complete AND logic operation in molecular domain. The AND gate can be converted to a NAND gate with the addition of a subtraction module.
     
\subsection{OR and NOR Gates}
\begin{figure}[!t]
	\centering
	\includegraphics[width=3.5in]{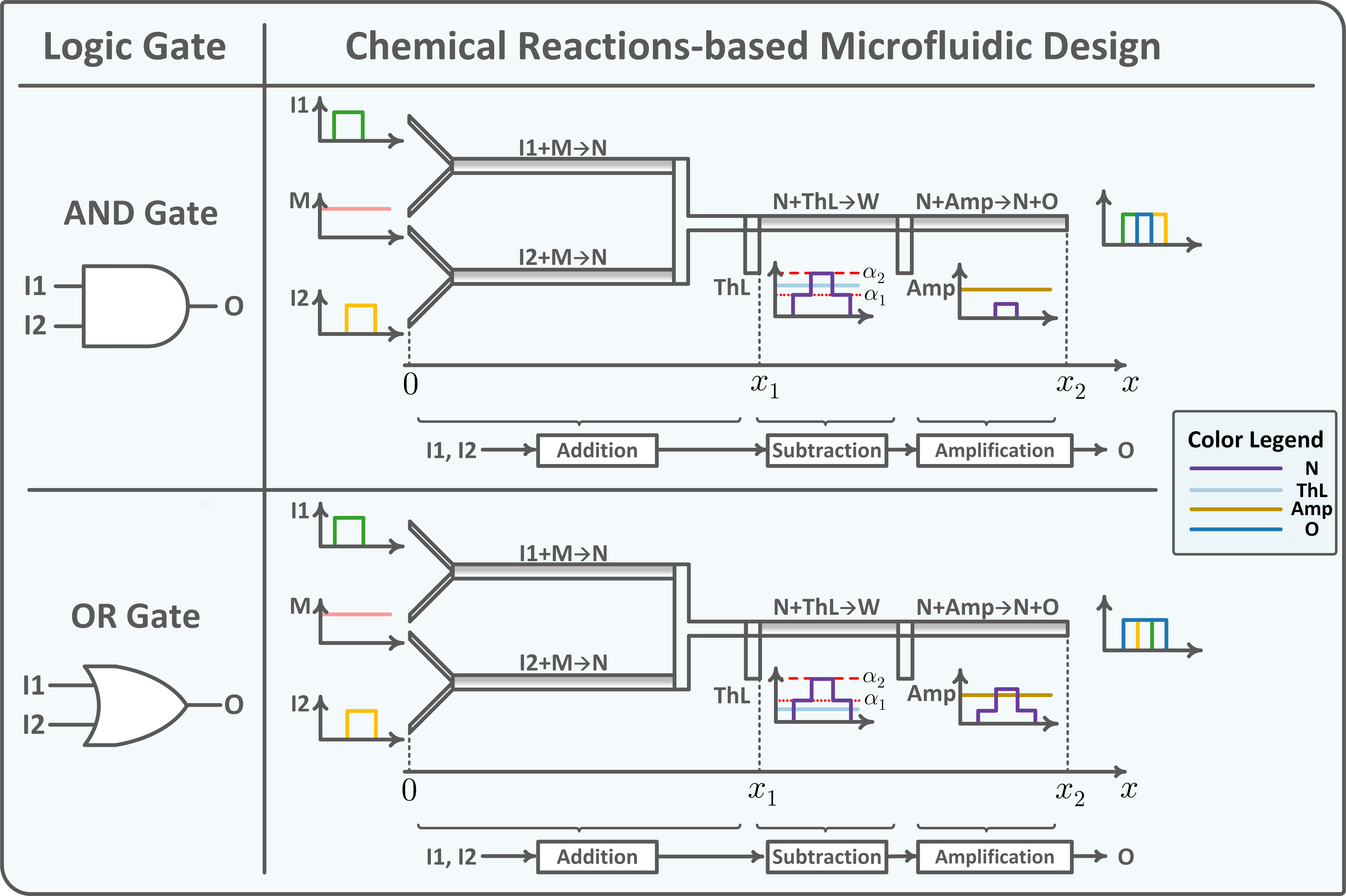}
	\caption{The chemical reactions-based microfluidic AND gate and OR gate. }
	\label{f_4}
\end{figure}
 
The OR gate can be designed using a similar geometry structure as the AND gate, as shown in Fig. \ref{f_4}.
Different from the AND gate, an OR gate generates a HIGH state for output species $O$ when one or both input species $I1$ and $I2$ are HIGH.
The only difference in design parameters between AND gate and OR gate is the injected concentration of species $ThL$ at the subtraction module. In theory, the concentration of species $N$ at location $x_1$ should be zero when both species $I1$ and $I2$ are LOW. However, this value is likely to be slightly larger than zero in practice. To mitigate this fluctuation, the output species $N$ generated by the addition module is required to be further processed by a subtraction module in which the concentration of species $ThL$ at location $x_1$ (i.e., $C_{ThL}$) should be larger than the fluctuation level and smaller than {$\alpha_1$} (i.e., the concentration of species $N$ at $x_1$ when only one input is HIGH). 
{When either one input is HIGH or both inputs are HIGH, the remaining concentrations of species $N$ after $N+ThL\to W$ have two different values, and an amplification module is used to ensure that these two values can lead to a generation of the same amount of output species $O$.}
The OR gate can also be converted to a NOR gate with a cascade of a
subtraction module.

\subsection{XOR Gate}
\begin{figure}[!t]
	\centering
	\includegraphics[width=3.5in]{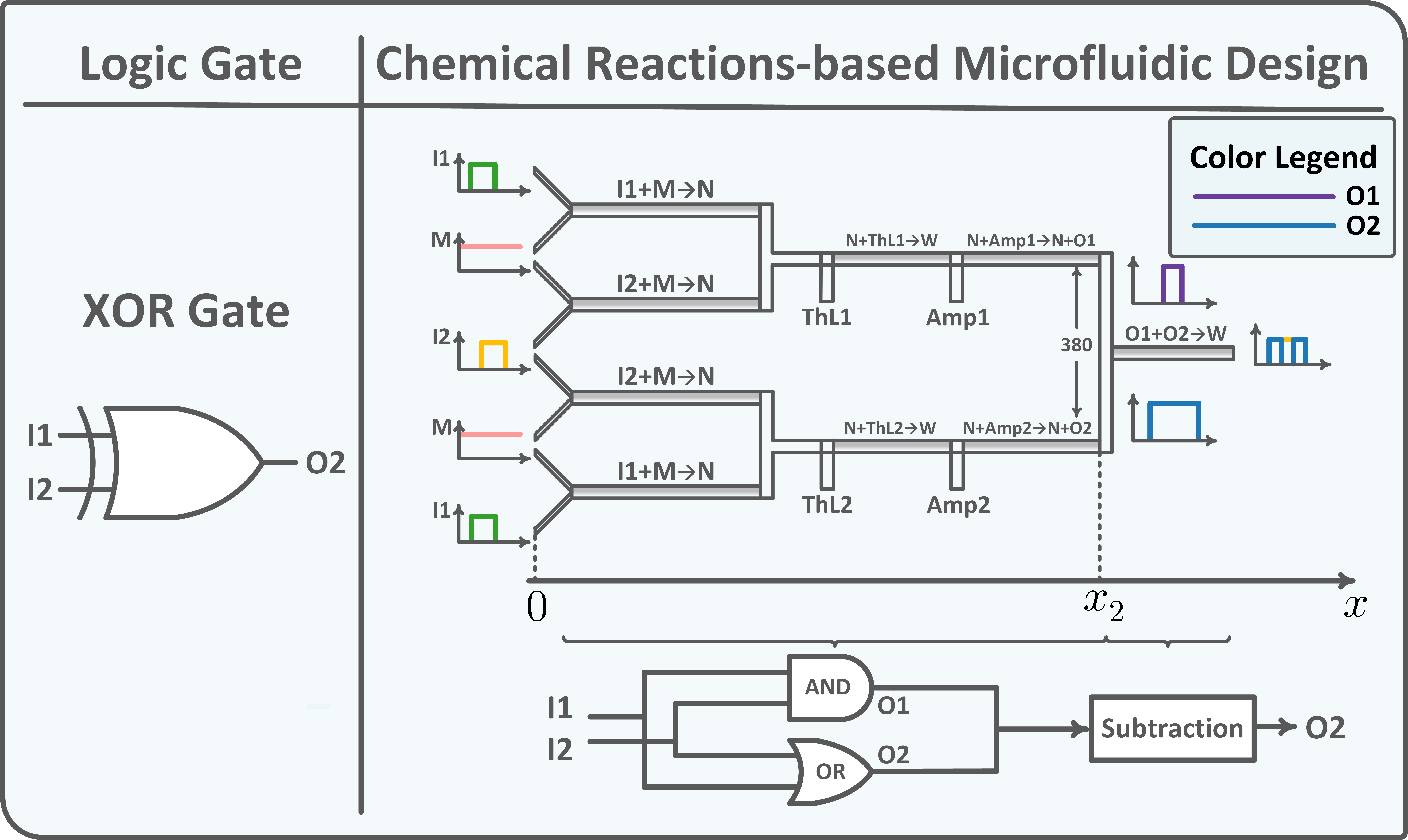}
	\caption{The chemical reactions-based microfluidic XOR gate.}
	\label{f_XOR}
\end{figure}
We design the XOR gate based on an AND gate, an OR gate, and a subtraction module. As shown in Fig. \ref{f_XOR}, input signals $I1$ and $I2$ first flow into the AND and OR gates that operate parallelly to generate species $O1$ and $O2$ at location $x_2$, respectively. Then, the generated species $O1$ and $O2$ enter a subtraction module to activate $O1+O2\to W$. In this reaction, species $O1$ is completely depleted by species $O2$ so that the remaining concentration of specie $O2$ only shows a HIGH state when either input species $I1$ or $I2$ is HIGH, thus achieving the XOR operation.

\subsection{Microfluidic Logic Gate Design Validation}
To examine the effectiveness of our designed logic gates, we simulate them in COMSOL. All the gates are constructed using the channel lengths marked in Figs. \ref{basic_modules} and \ref{f_XOR} along with width $20$ $\mu$m and depth $10$ $\mu$m. As an exception, the width for Y junction is $10$ $\mu$m. 
We set the diffusion coefficient and mean injection velocity for each species as $10^{-8}$~m$^2$/s and $0.75$~cm/s, and the rate constant for each reaction as $5000$~m$^3$/(mol$\cdot$s). For all the gates, we set injected concentrations of species $I1$, $I2$, and $M$ as $C_{I1}(t)=8[u(t-1)-u(t-3)]$, $C_{I2}(t)=8[u(t-2)-u(t-4)]$, and $C_M(t)=8u(t)$ with Heaviside step function $u(t)$. 
For other species, we set: $C_{ThL}(t)=6u(t)$ and $C_{Amp}(t)=4u(t)$ for AND gate; $C_{ThL}(t)={2u(t)}$ and $C_{Amp}(t)=4u(t)$ for OR gate; $C_{ThL1}(t)=6u(t)$, $C_{Amp1}(t)=4u(t)$, $C_{ThL2}(t)={2u(t)}$, and $C_{Amp2}(t)=4u(t)$ for XOR gate. The concentration unit is mol/m$^3$.

\begin{figure}%
	\centering
	\includegraphics[width=3.5in]{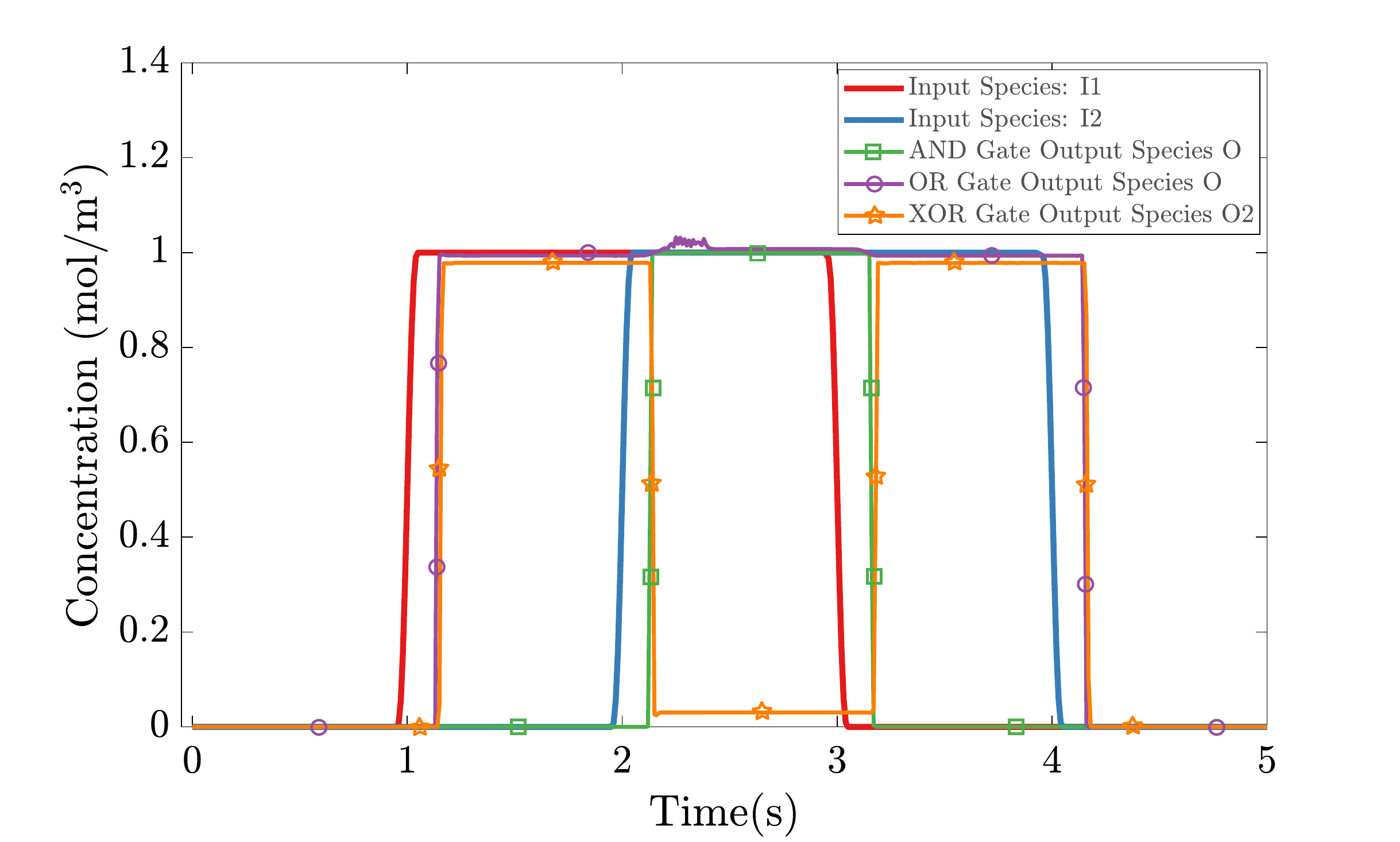}
	\caption[]{The normalized concentrations of input species $I1$ and $I2$ and the circuits' outputs. }
	\label{f_sim}%
\end{figure}

The simulation results are plotted in Fig. \ref{f_sim}. We observe that our designed microfluidic circuits show desired behavior as their corresponding electric circuits in terms of logic gate operation functionalities, which demonstrates that our designed circuits are feasible and effective for digital signal processing in molecular domain. 

\section{Challenges and Potential Solutions}
\label{challenge}

The realization of signal processing functions via microfluidic circuits shows significant advantages in reusability and modularity. 
On the one hand, the reusability is reflected on that the subtraction-amplification module and the AND-OR gate share the same microfluidic geometry structure but with different reactions and species concentrations, as shown in Figs. \ref{basic_modules} and \ref{f_4}.
Hence, a microfluidic structure can perform different functions by using different design parameters, which reduces implementation cost and enables a separation of function design from device manufacture. On the other hand, the modularity is embodied in the construction of different logic gates. As illustrated in Figs. \ref{f_4} and \ref{f_XOR}, our designed microfluidic logic gates can be constructed via the combinations of three microfluidic modules at Level 3, which is similar to Legos that a construction of vehicles, buildings, or working robots are built merely via interlocking plastic bricks.

\begin{figure}[!t]
	\centering
	\includegraphics[width=3.6in]{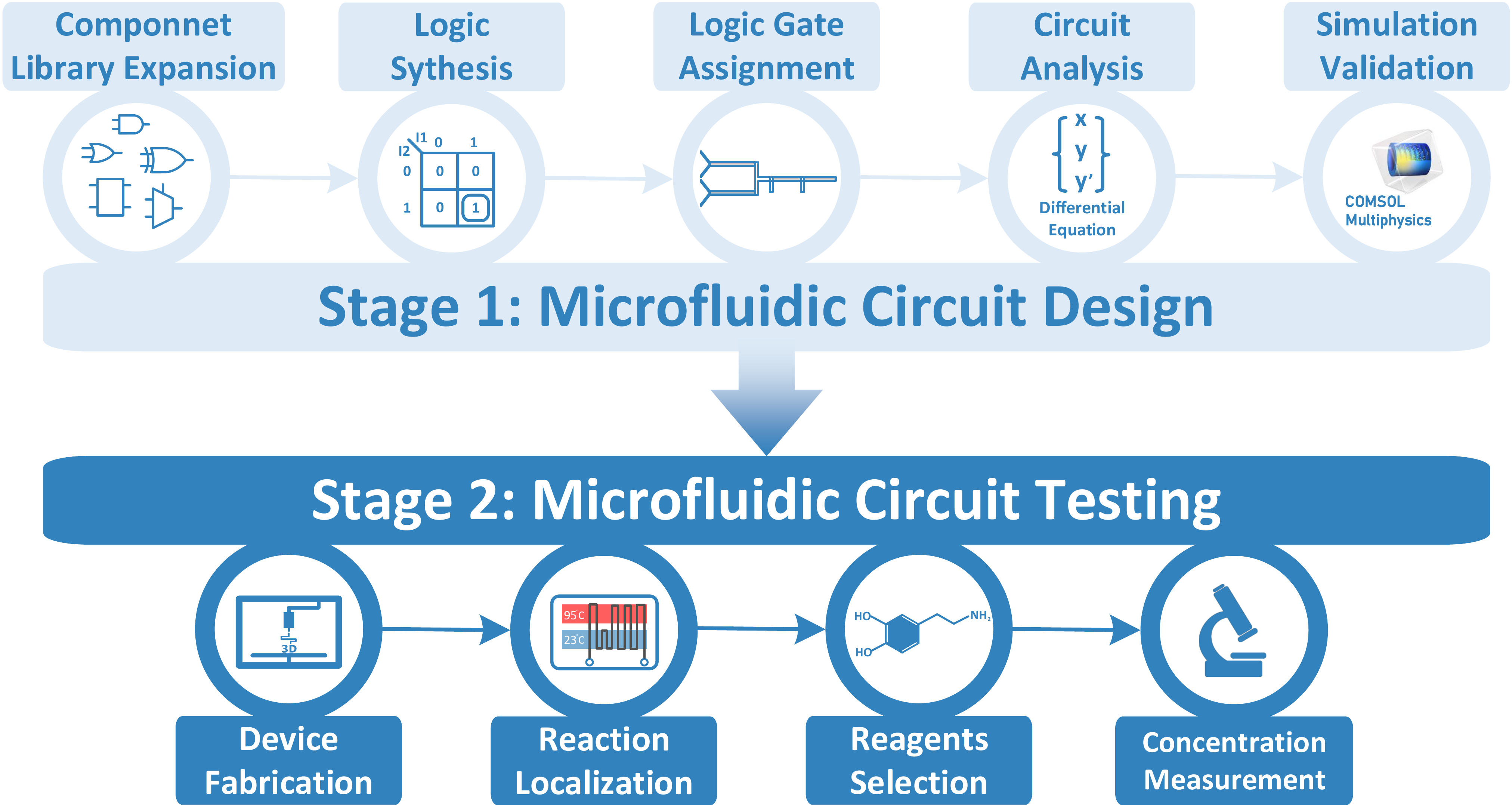}
	\caption{A roadmap for the development of microfluidic circuits with complex signal processing functions.}
	\label{f_flow}
\end{figure}

Due to reusability and modularity, more complex signal processing circuits at Level 5 are envisioned to be built through combinations of logic gates at Level 4. 
In order to clarify the steps moving from Levels 1 to 5,
we illustrate a roadmap 
in Fig. \ref{f_flow}, which includes microfluidic circuit design and testing stages. In the following, we highlight the main challenges in each stage and identify corresponding potential solutions.

\subsection{Microfluidic Circuit Design}
The complexity of a signal processing function largely depends on the number of available gates; thus, the first step in the design stage is to expand logic gate library. Second, logic synthesis should be performed to identify the circuit diagram for a specified operation, 
which provides a basis for the followed gate assignment to choose correct gates. Then, functional connecting gates should be theoretical analyzed and verified by simulation results.

\subsubsection{Component Library Expansion}
Compared with the number of electronic logic gates in the literature, the number of microfluidic logic gates is still limited. Thus, it is essential to design more microfluidic logic gates, such as multiplexers and decoders, and expand the component library to allow for more complex signal processing. As the microfluidic logic gates at Level 4 are built from the microfluidic modules at Level 3, the library expansion includes designing or introducing more components at lower levels. This includes but not limited to introduce serpentine and herringbone-like geometry for Level 1, design biological inspired chemical reactions for Level 2, and basic arithmetic operations (e.g., half adder) for Level 3.

It is noted that our designed logic gates are all combinational circuits, where circuits' outputs only depend on current input signals. However, the realization of many signal processing functions can be based on sequential logic circuits, where circuits' outputs depend on both current and previous input signals. 
In microfluidic circuits, how to retain flowing chemical information to achieve sequential circuits is a big challenge.

\subsubsection{Logic Synthesis}
Logic synthesis produces a circuit diagram with available logic gates to perform a specified operation. A challenge here is how to determine the types of gate for a desired operation. 
This procedure is usually not straightforward and explicit, 
but we can provide valuable insights from some cases to facilitate logic synthesis. For example, to achieve the $n$th order CSK modulation, we can translate this operation into a control problem, which can be implemented by a $n:2^n$ decoder so that one concentration level is transmitted from $2^n$ levels in terms of the combination of $n$ input bits \cite{bi2020microfluidic}. In addition, it would be helpful to use Boolean algebra and Karnaugh maps to derive the simplest Boolean equations that reduce a circuit redundancy and lead to the simplest set of logic gates. 


\subsubsection{Logic Gate Assignment}

After determining a circuit diagram and the required gate types via logic synthesis, the next step is to select available gates from logic library to be used in the circuit. We note that even for a \textit{single} gate in a \textit{whole} circuit, it is likely to have different microfluidic designs, which introduce a challenge to identify the optimal gate combination for the whole circuit. A straightforward method 
is to permute all possible designs for each gate and identify the optimal combination of the whole circuit. 
However, this approach becomes intractable with the increase of library size and circuit scale. To address this challenge, attention needs to be given to design an assignment algorithm.
Furthermore, quantified metrics should be developed to analyze circuit complexity, such as the total size of a circuit, the total number of species and chemical reactions, and the speed to finish specified signal processing tasks.
\subsubsection{Circuit Analysis}
An integral part of microfluidic circuit design is how to derive circuits' outputs and theoretically analyze circuit performance, such as the channel noise characterization and channel capacity calculation for a circuit with communication functionalities. By doing so, it would reveal the dependency of circuit performance on design parameters, and provide feedback for the circuit design.



To derive circuits' outputs, we can rely on the impulse response of a microfluidic channel so that output signals can be written as the convolution of circuit inputs and a cascade of the 
impulse response of each channel. Unfortunately, there are many challenges in merely deriving the impulse response of an individual channel, and we highlight two main challenges as follows:
\begin{itemize}
	\item The turbulent flow: It has been revealed that flows in microfluidic channels are mostly laminar, 
	and a 3D partial differential equation (PDE) can be approximated by a 1D PDE when laminar flows fall into the dispersion regime \cite{9090868}. However, microfluidics may still exhibit turbulent feature, which can be caused by unintended physical barriers owing to the imperfectness of device fabrication, or intended physical barriers (e.g., herringbone-like geometry) designed to facilitate species mixing. It is noted that turbulent flow is always 3D, and this characteristic hinders the simplification of 3D PDEs.
	\item The coupling of convection, diffusion, and reaction: Chemical reactions occur during molecular movement such that chemical reactions are fully coupled with convection and diffusion process, leading to the non-linearity of the convection-diffusion-reaction equation. 
	To deal with this problem, we can rely on operator splitting method \cite{bi2020microfluidic} that splits the original PDE into several easily solved subproblems, and perturbation method \cite{abin2020analytical} that assumes the analytical solution in the form of an infinite series of functions.
\end{itemize}

\subsubsection{Simulation Validation}
Although theoretical analysis can predict circuits' outputs, it is necessary to validate well-designed circuits via appropriate simulation tools before prototyping. For microfluidic circuits, finite element simulation is the most suitable simulation approach. In finite element simulations, a microfluidic channel is partitioned into small meshes with geometrically simple shapes, and molecular concentrations in each mesh are updated over a sequence of time steps according to the corresponding PDEs. Not surprisingly, the mesh settings, such as 
the size, the density, and the number of meshes, impose an impact on the time and the amount of memory required to compute a design, and finally influence the accuracy of the solution. With an increase of circuit scale, the default meshing sequences provided by simulation tools may not guarantee an accurate solution within a short time, which encourages engineers to build a customized mesh that is best-suited for their particular models. Therefore, how to quickly find a customized and an optimal mesh remains a major challenge.

\subsection{Microfluidic Circuit Testing}


After verifying a well-defined microfluidic circuit, the next stage is to build a circuit prototype, including circuit fabrication, chemical reaction localization, and reagents selection. Moreover, detection techniques should also be selected to visualize the circuit's outputs.

\subsubsection{Device Fabrication}
Over the past two decades, a majority of microfluidic devices have been built in poly(dimethylsiloxane) (PDMS) by soft lithography due to some key properties, including biocompatibility, transparency, and water-impermeability. 
However, PDMS fabrication usually involves substantial human labor and layered molding, which tends to hinder the dissemination of PDMS devices outside of research labs and the production of complex 3D devices. 
Fortunately, the rapidly developed 3D-printing is a promising technique for microfluidic fabrication with advantages of automation and assembly-free 3D fabrication. 
Moreover, the thermoplastic materials-based fabrication method is also an alternative. Thermoplastic materials are not only more robust and easier to manufacture than traditional materials but can also achieve a rapid and low cost fabrication.

\subsubsection{Reaction Localization}
How to confine a chemical reaction within a region to achieve desired circuit behavior is a critical challenge. According to the \textit{Arrhenius} equation that reveals a dependency of rate constant on temperature, temperature control provides an opportunity to address this challenge. By means of cooling convection-diffusion channels while heating reaction channels, it would allow us to keep convection-diffusion channels thermally isolated from reaction channels, which ensures pre-mixed reactants do not react until they reach heated reaction regions \cite{yen2005microfabricated}.

\subsubsection{Reagents Selection}
In the testing stage, a key step is to choose appropriate reagents to map to the species in chemical reactions.
The selection of reagents should consider the following aspects:
\begin{itemize}
	\item The selected reagents and chemical reaction products must be non-toxic to human body and environment.
	\item The interactions among selected reagents, reaction products, and channel materials should be studied so as to prevent any side reactions.
	\item The outputs of a logic gate are expected to be the inputs of the cascaded gate. By doing so, the interconnection of logic gates can be automated.
	\item The disposition or recycle of remaining solutions for future use should be taken into account.
\end{itemize}
With the above requirements, the number of satisfied reagents may be limited, which imposes a restriction on circuit sizes and functions.

\subsubsection{Concentration Measurement}
The outputs of microfluidic circuits 
should be analyzed by suitable detection techniques, which presents another major challenge. There are two main factors that affect the choice of concentration detection method: the sensitivity to output species and the scalability to smaller dimensions. Among the common detection techniques, 
the optical-based technique is prone to satisfy the requirements of sensitivity and scalability. In particular, the ``on-chip'' detection mode, where some optical components have been fully integrated with or fabricated together with microfluidic devices, brings new opportunities for concentration detection. This mode strongly ties optics and microfluidics together, which exhibits great advantages in high interaction efficiency compared with traditional ``off-chip'' optics configuration where detection units are separated from microfluidic devices \cite{yang2018micro}.

\section{Conclusion}   

The main hindrance for molecular communication (MC) applications stands in the lack of nano/micro-devices able to process chemical concentration signals in biochemical environment. In this article, we proposed the vision of using chemical reactions-based microfluidic circuits to perform signal processing functions. To manage the complexity of microfluidic processing systems, we described a microfluidic circuit from five levels of abstraction:  1) Molecular Propagation; 2) Chemical Transformation; 3) Microfluidic Modules; 4) Microfluidic Logic Gates; and 5) Microfluidic Circuits. We first introduced channel geometry in Level 1 and chemical reactions in Level 2. We then presented the designs of microfluidic modules in Level 3 and microfluidic logic gates in Level 4. Importantly, these designs demonstrated significant advantages in reusability and modularity, and this motivated us to design the microfluidic circuits in Level 5 with complex signal processing functions based on the designs in Levels 1-4. We identified the challenges in designing and testing complex microfluidic circuits and proposed corresponding potential solutions. This article serves to inspire research to design, analyze, and test novel chemical reactions-based microfluidic circuits that perform signal processing functions in molecular domain, in order to support the advancement of MC-enabled applications.



%

\appendices




\ifCLASSOPTIONcaptionsoff
  \newpage
\fi



%
%

\vskip -2\baselineskip plus -1fil

\begin{IEEEbiographynophoto}{Dadi Bi}
	is currently pursuing the Ph.D. degree with the Department of Engineering, King's College London, U.K. His research interests include molecular communication and microfluidics.
\end{IEEEbiographynophoto}

\vskip -2\baselineskip plus -1fil

\begin{IEEEbiographynophoto}{Yansha Deng}
 is currently a Lecturer (Assistant Professor) with the Department of Engineering, King's College London, U.K. Her research interests include molecular communication, machine learning, and 5G wireless networks. 
\end{IEEEbiographynophoto}

%




\end{document}